\input{aipcheck}

\documentclass[
    ,final            
  ]
  {aipproc}

\layoutstyle{6x9}

\begin{document}

\title{Topological effects on string vacua}


\author{Oscar Loaiza-Brito}{
  address={Departamento de F\'isica, Divisi\'on de Ciencias e Ingenier\'ias, Campus Le\'on, Universidad de Guanajuato, P.O. Box E-143, Le\'on, Guanajuato, M\'exico}
}

\begin{abstract}
 We review some topological effects on the construction of string flux-vacua. Specifically we study the effects of brane-flux transitions on  the stability of D-branes on a generalized tori compactificaction, the transition that a black hole suffers in a background threaded with fluxes  and the connections among some Minkowsky vacua solutions.
\end{abstract}

\maketitle


\section{Introduction}
During the last decade,  flux string compactifications  have become an important setup on the construction of realistic string vacua \cite{Grana:2005jc}. The main reason for that lies on the possibility to construct a superpotential depending on the compactificaction moduli which in turn yields their stabilization. However, the presence of such fluxes impose some stringent constraints on the models involving wrapping D-branes on internal cycles. An important consequence is the realization of a topological transition between a brane configuration into one involving fluxes. 

Wrapping a D-brane on a submanifold supporting  NS-NS flux flux makes the D-brane,  Freed-Witten (FW) anomalous \cite{Freed:1999vc}. This anomaly in the corresponding string worldsheet, can be understood in the following way. 
Consider a $D(p+2)$-brane wrapping a submanifold ${\cal W}_{p+3}$.  If there exists a NS-NS 3-form $H_3$ with its three legs on the worldvolume of the $D$-brane (a non-vanishing pullback of $H_3$ on ${\cal W}_{p+3}$), a monopole charge is induced through the action term $\int_{\cal W} A_{p}\wedge H_3$, where $A_{p}$ is the dual gauge potential on ${\cal W}_{p+3}$. The corresponding equations of motion are not fulfilled unless the source for the magnetic charge is added. This is accomplished by considering an extra $D$-brane ending at the worldvolume of the FW-anomalous brane. Altogether, the system is well defined and consistent.

Hence, in the presence of NS-NS flux, $D$-branes cannot wrap any submanifold. A consistent brane configuration should involve a net of submanifolds on which branes of different  dimensions are wrapped. Specifically, if a $D(p+2)$-brane wraps a  submanifold ${\cal W}_{p+3}$ on which the pullback of $H_3$ is different from zero,  an extra $Dp$-brane wrapping a  $(p-1)$-submanifold of ${\cal W}_{p+3}$ and extending on one coordinate on the transversal space to ${\cal W}_{p+3}$ is required. Being the submanifolds adequate to define spinors and their coupling with gauge fields, this configuration guarantees the absence of  inconsistencies.

There are however,  D-branes which in spite of being FW anomaly-free, are nevertheless unstable. This is physically interpreted as a topological transformation between D-branes and fluxes through the appearance of instantonic branes \cite{Maldacena:2001xj}. Formally, the transition is described by the Atiyah-Hirzebruch Spectral Sequence which roughly speaking, connects cohomology with twisted K-theory. A more pedestrian way to visualize the transition requires a coordinate rotation in which the extra brane added to cancel the FW anomaly intersects the anomalous brane in a timelike coordinate. Hence, an apparent  stable D-brane in a background threaded with NS-NS flux,  moves in time an encounters an instantonic  brane. The encounter makes the smaller brane unstable to decay into a configuration of fluxes consisting on the coupling between the NS-NS flux and the magnetic field strength of the instantonic brane.

\section{Freed-Witten anomaly in generalized tori compactifications}
A six-dimensional torus threaded with NS-NS flux $H_{abc}$ is mapped under T-duality (on coordinate $x^a$) into a "twisted torus" which is a nilpotent manifold  with a structure constant $f_{bc}^a$. The twisted torus is the quotient space constructed by the relations  $(x^a, x^b, x^c, ...)\sim (x^a+1, x^b, x^c,...)\sim (x^a, x^b+1, x^c, ...)\sim (x^a +f^a_{bc}x^b, x^c+1, ...)$ in which there is not NS-NS flux \cite{Wecht:2007wu}.  In this sense,  all the effects of having a NS-NS background flux are mapped into effects produced by the new geometry represented by $f^a_{bc}$ which is an integer number referred to as  {\it metric flux}. 

In particular, 
a D-brane wrapping a cycle which ends at an instantonic brane supporting some NS-NS flux, is mapped under T-duality into a D-brane wrapping a torsion cycle in the twisted torus \cite{Marchesano:2006ns}.
According to the physical interpretation given by MMS, this means  that $N$ D$p$-branes wrapping a torsion cycle will decay into fluxes by encountering an instantonic D$p$-brane wrapped in a spatial $(p+1)$-cycle supporting the metric fluxes. The remnant fluxes are given by the coupling between metric fluxes and the magnetic field strength related to the instantonic brane \cite{LoaizaBrito:2006se}. This allows D$p$-branes to transform into $f^a_{bc}F_{a\mu_1\dots \mu_{3-p}}$.

However,  under the above T-duality map, the toroidal  K\"ahler  closed form $J$  ($dJ=0$) is mapped into a non-closed form ($dJ\neq 0$). Hence by turning on an extra complex flux, composed by the NS-flux and the non-closed K\"ahler form,  one can see that the flux
$(H_3+f J)$, is the source of instabilities for some D-branes wrapping internal cycles in the twisted torus.

The corresponding physical interpretation is as follows. Some D-branes\footnote{Those representing trivial cocycles under the map $\overline{d}_3= (H_3+fJ)\wedge$} would decay into a configuration of fluxes by encountering an instantonic D$(p+2)$-brane wrapped in a cycle which in turn supports the complex flux $H_3+f J$. The remnant is the coupling between the complex three-form and the magnetic field strength for the instantonic brane. Hence some D$p$-branes, would transform\footnote{If the transition is energetically favorable.}  topologically into $(H_3+f J)\wedge\ast F_{p+4}$. 

Interesting enough, one can see  that in some cases, as for three-cycles in the above simple twisted torus, we can construct
a chain of instantonic branes in which the flux $\omega \ast F_{p+2}$ transforms into $(H_3+\omega J)\wedge\ast F_{p+4}$, relating NS-NS fluxes with RR ones and vice versa. An explicit example \cite{LoaizaBrito:2006se} shows that for a type IIA compactification on a twisted six-torus threaded by a metric flux in coordinates 456 and a NS-flux compatible with supersymmetric conditions, there are only 6 cycles out of 16, where a space-filling D6-brane can be wrapped.

Moreover, by studying the case of the twisted six-torus orbifolded by ${Z}_2\times {Z}_2$, one finds that all D6-branes wrapping the invariant (untwisted) cycles, are either unstable to decay into fluxes or Freed-Witten anomalous. In cases like these, phenomenological models are not protected against instabilities unless $H_3+f J=0$. It is important to say that we have considered only forms and cycles invariant under the action of the above discrete group. It would be interesting to extend this study to $Z_2$-twisted forms as well.



%


\section{Black Holes and fluxes}
A four-dimensional black hole in ${\cal N}=2$ supersymmetric background can be constructed by wrapping several D3-branes on internal 3-cycles. The magnetic and electric charges carried by the black hole are those carried by the D3-branes under the potential Ramond-Ramond (RR) 4-form $C_4$ \cite{Suzuki:1995rt}.

In a flux compactification, the underlying geometry backreacts due to the fact that the fluxes gravitate and it is no so clear  weather a black hole solution consisting on wrapping D3-branes would be possible. In \cite{Danielsson:2006jg} the authors study how the supersymmetric vacua solutions in the presence of fluxes are altered by the black hole, concluding that in a generic description, they are not. 

In \cite{LoaizaBrito:2007kz}  
it was shown that, under the presence of the non trivial NS-NS three-form flux~$H_3$ in the internal space, the four dimensional black hole described by D3-branes  may disappear via the topological process that transforms branes into fluxes.
A NS-NS flux induces the FW anomaly on an instantonic D5-brane which must be canceled by a D3-brane ending on it.
Applying this fact to the system of the D3-branes wrapped on an internal Calabi-Yau cycle under the presence of extra NS-NS flux~$H_3$, one sees that the four-dimensional black hole would suffer from the same topological transition.

        Under the four-dimensional perspective, the black hole would disappear leaving as a remnant a RR three-form flux~$F_3$ localized in the uncompactified four spacetime dimensions. 
        In order to compute the black hole quantities, it is assumed that the black hole represents a BPS state even in the presence of the NS-NS flux~$H_3$, namely, that there is not interplay between the effective scalar potential induced by the presence of the fluxes and the extremal black hole except for the transition drived by the instantonic brane. This means 
 that the BPS states in the usual ${\cal N}=2$ four-dimensional supergravity are preserved in the ${\cal N}=2$ gauged supergravity.
The charge (and mass) carried by the black hole before the topological transition driven by the instantonic D5-brane is afterwards carried by the coupling between the NS-NS flux $H_3$ and another RR three-form flux $F_3$ emanating from the D5-brane, implying that even the extremal black hole suffers from the transition that exchanges topologically different configurations which carry the same charge and mass.

Recently, it has been conjectured that an ensemble of the different solutions without horizon corresponds to a black hole~\cite{Mathur:2005zp}. 
According to it, the horizon radius is nothing but the size of the region where the solutions in the ensemble differs each other.
%
%

There are many open questions.
First of all, the already mentioned topological transformation was explored some years ago in the context of the conifold approximation~\cite{Greene:1995hu}.
In this scheme, the moduli space describing the geometry of the conifold on which a D3-brane is wrapped changes after a topological transformation under which the three-cycle of the confiold shrinks to zero and blow up into a two-cycle. Particularly, the complex structure moduli were exchanged into the K\"ahler moduli. It would be very interesting to go further and study what happens with the Calabi-Yau moduli under the presence of an instantonic brane.

\section{Minkowski vacua transitions}
Canceling the FW anomaly by the addition of extra branes makes the superpotential $\cal{W}$
to be non-invariant, implying the presence of connections between different Minkowski vacuum solutions from F-flat conditions. Although brane-flux transitions leave the tadpole condition invariant they perform a change in  $\cal{W}$ which depends on RR fluxes. This essentially follows from the fact that there is a remnant  of one unit of RR flux attributed to the instantonic brane. By considering the S-dual case as well,  brane-flux transition is driven also by instantonic NS5-branes.\\

Vacuum solutions of type IIB flux compactifications on a factorizable six-dimensional torus in a simple orientifold plane, can be written in terms of polynomials $P_1$ and $P_2$ which share a common factor $P$. All polynomials depend on the complex structure $U=i\rho$ and on the units of RR and NS-NS flux. Before the transition, complex structure $\rho$ is stabilized at the commun root $\rho_0$ of $P(\rho)$, and the dilaton $S(\rho)$ fulfills the equation $P_1'(\rho)-SP_2'(\rho)=0$. After the transition there are three general cases \cite{HerreraSuarez:2009ne}:

1.  $P_1(\rho)=P(\rho)Q_1(\rho)\rightarrow \hat{P}_1(\rho)=P(\rho)\hat{Q}_1(\tilde{f}_1\rho+\tilde{g}_1)$, where $\tilde{f}$ and $\tilde{g}$ are functions of the modified RR fluxes. The complex structure is invariant (there is still a common root between $P_2$ and $\hat{P}_1$), but the value at which the dilaton is stabilized suffers a change. The prototypical example for this case takes a transformation on $P_1(\rho)$ to $\kappa P_1(\rho)$ from which one gets that the complex structure is stabilized at the same value $\rho_0$.  However, from the F-flat equation $\partial_\rho{\cal{W}}_H=0$ we get that the dilaton $S$ must fullfill the equation
$\kappa P_1'(\rho)- S P_2'(\rho)=0$,
which is true for $S=\kappa \rho_0$.

2. The transition changes the polynomial such that there is  non-common factor $P(\rho)$. The new polynomial  reads $\hat{P}_1(\rho)=\hat{P}(\rho)\hat{Q}_1(\rho)$ with $\hat{P}(\rho) =0$ at $\rho=\rho_*(\neq \rho_0)$. Hence, at $\rho=\rho_\ast$  and at $\rho=\rho_0$, ${\cal{W}}_H\neq 0$ from which we see that supersymmetry is broken through all moduli. At both points, the scalar potential is positive (since the model is no-scale) but an extended and more detailed analysis is required to elucidate if the scalar potential is minimum. In the case in which the scalar potential is not minimum at those points, we can say that brane-flux transitions connect the supersymmetric configuration into some unstable excitations around the basis point.

3.  By considering transitions mediated by NS5-branes,
we find an example in which the transition changes the polynomial $P_1(\rho)$ and $P_2(\rho)$  such that there is still a common root since they share a common factor.  However, this common factor $Q(\rho)$ has also changed with respect to $P(\rho)$ gathered from the original set of fluxes. The new polynomials are  $P_1(\rho)= Q(\rho)(\tilde{f}_1\rho +\tilde{g}_1)$ and $P_2(\rho)=Q(\rho)(\tilde{f}_2\rho + \tilde{g}_2)$ with $Q(\rho_\ast)=0$ and $\rho_\ast\neq\rho_0$. In this case, the transition connects two different supersymmetric solutions which differ not only by a rescaling of fluxes. Therefore, the nucleation of instantonic NS5-branes
allows us to connect many different vacuum solutions which otherwise would be disconnected. The S-dual version of the brane-flux transition must  notably reduce the size of vacuum solutions.\\

These results do not involve the presence of non-geometric fluxes. Their incorporation establishes novel ways to increase the value of RR and NS-NS fluxes.  For 
simple factorizable torus compactifications, the following is observed:

1. For the cases in which the solution to SUSY equations allows a non vanishing complex structure $U$ with $Re(U)>0$, the configuration of fluxes does not satisfy the required constraints for the transition to happen. This mean that brane-flux transition is forbidden for these cases, isolating the vacuum solution to connect to others. Notice that these solutions are interesting from the phenomenology point of view and are protected to move to another vacuum via an instantonic brane mediation.

2. For the cases in which one gets a pure imaginary complex structure $U$, the set of fluxes satisfy the constraints which allow the transition to occur. Different vacuum solutions, all of them sharing the property that $Re(U)=0$, are connected through a chain of  instantonic branes.\\

It is important to  remark that such results are obtained under the supergravity limit, which might be not valid for degenerated torus as in our examples where $Re(U)=0$. 
It is also clear that a deeper  study on different vacuum solutions is required.


\begin{theacknowledgments}
  I thank the organizers for inviting me to give a talk, specially to A. G\"uijosa and E. C\'aceres for kind support. The work was partially supported by Conacyt  under the project  No. 60209. 
\end{theacknowledgments}



\bibliographystyle{aipproc}   

\bibliography{sample}

\hyphenation{Post-Script Sprin-ger}
\begin{thebibliography}{12}
\expandafter\ifx\csname natexlab\endcsname\relax\def\natexlab#1{#1}\fi
\providecommand{\enquote}[1]{``#1''}
\expandafter\ifx\csname url\endcsname\relax
  \def\url#1{\texttt{#1}}\fi
\expandafter\ifx\csname urlprefix\endcsname\relax\def\urlprefix{URL }\fi
\providecommand{\eprint}[2][]{\url{#2}}

\bibitem[Grana(2006)]{Grana:2005jc}
M.~Grana, \emph{Phys. Rept.} \textbf{423}, 91--158 (2006),
  \eprint{hep-th/0509003}.

\bibitem[Freed and Witten(1999)]{Freed:1999vc}
D.~S. Freed, and E.~Witten  (1999), \eprint{hep-th/9907189}.

\bibitem[Maldacena et~al.(2001)]{Maldacena:2001xj}
J.~M. Maldacena, G.~W. Moore, and N.~Seiberg, \emph{JHEP} \textbf{11}, 062
  (2001), \eprint{hep-th/0108100}.

\bibitem[Wecht(2007)]{Wecht:2007wu}
B.~Wecht, \emph{Class. Quant. Grav.} \textbf{24}, S773--S794 (2007),
  \eprint{0708.3984}.

\bibitem[Marchesano(2006)]{Marchesano:2006ns}
F.~Marchesano, \emph{JHEP} \textbf{05}, 019 (2006), \eprint{hep-th/0603210}.

\bibitem[Loaiza-Brito(2007)]{LoaizaBrito:2006se}
O.~Loaiza-Brito, \emph{Phys. Rev.} \textbf{D76}, 106015 (2007),
  \eprint{hep-th/0612088}.

\bibitem[Suzuki(1996)]{Suzuki:1995rt}
H.~Suzuki, \emph{Mod. Phys. Lett.} \textbf{A11}, 623--630 (1996),
  \eprint{hep-th/9508001}.

\bibitem[Danielsson et~al.(2006)]{Danielsson:2006jg}
U.~H. Danielsson, N.~Johansson, and M.~Larfors, \emph{JHEP} \textbf{09}, 069
  (2006), \eprint{hep-th/0605106}.

\bibitem[Loaiza-Brito and Oda(2007)]{LoaizaBrito:2007kz}
O.~Loaiza-Brito, and K.-y. Oda, \emph{JHEP} \textbf{08}, 002 (2007),
  \eprint{hep-th/0703033}.

\bibitem[Mathur(2005)]{Mathur:2005zp}
S.~D. Mathur, \emph{Fortsch. Phys.} \textbf{53}, 793--827 (2005),
  \eprint{hep-th/0502050}.

\bibitem[Greene et~al.(1995)]{Greene:1995hu}
B.~R. Greene, D.~R. Morrison, and A.~Strominger, \emph{Nucl. Phys.}
  \textbf{B451}, 109--120 (1995), \eprint{hep-th/9504145}.

\bibitem[Herrera-Suarez and Loaiza-Brito(2010)]{HerreraSuarez:2009ne}
W.~Herrera-Suarez, and O.~Loaiza-Brito, \emph{Phys. Rev.} \textbf{D81}, 046002
  (2010), \eprint{0906.0608}.

\end{thebibliography}

\IfFileExists{\jobname.bbl}{}
 {\typeout{}
  \typeout{******************************************}
  \typeout{** Please run "bibtex \jobname" to optain}
  \typeout{** the bibliography and then re-run LaTeX}
  \typeout{** twice to fix the references!}
  \typeout{******************************************}
  \typeout{}
 }

\end{document}